\documentclass[10pt,journal,compsoc]{IEEEtran}

\ifCLASSOPTIONcompsoc
  \usepackage[nocompress]{cite}
\else
  \usepackage{cite}
\fi

\ifCLASSINFOpdf
\else
\fi

\usepackage{bm}
\usepackage{enumerate}
\usepackage{enumitem}
\usepackage{makecell}
\usepackage{multirow}
\usepackage{stfloats}
\usepackage{amsmath}
\usepackage{color}
\usepackage{amsfonts}
\usepackage{mathtools} 
\usepackage{amssymb}
\usepackage{verbatim}
\usepackage[normalem]{ulem}
\useunder{\uline}{\ul}{}
\usepackage{booktabs}
\usepackage{dsfont}
\usepackage[table]{xcolor}

\usepackage{algorithm}
\usepackage{algorithmic}
\usepackage{enumerate}

\allowdisplaybreaks[3]

\begin{document}

\title{Prompt-based Multi-interest Learning Method for Sequential Recommendation}

\author{Xue~Dong,
        Xuemeng~Song,~\IEEEmembership{Senior Member,~IEEE},
        Tongliang Liu,~\IEEEmembership{Senior Member,~IEEE},
        and~Weili Guan
\IEEEcompsocitemizethanks{\IEEEcompsocthanksitem X. Dong is with the School of Software, Tsinghua University, Beijing 100084, China. E-mail: dongxue.sdu@gmail.com.
\IEEEcompsocthanksitem X. Song is with the School of Computer Science and Technology, Shandong University, Qingdao 266237, China. E-mail: sxmustc@gmail.com.
\IEEEcompsocthanksitem T. Liu is with Sydney AI Centre, the University of Sydney, 6 Cleveland St, Darlington, NSW 2008, Australia. Email: tongliang.liu@sydney.edu.au.
\IEEEcompsocthanksitem W. Guan is with the Faculty of Information Technology, Monash University (Clayton Campus), Australia. E-mail: honeyguan@gmail.com.
}
\thanks{X. Song is the corresponding author.}}

\markboth{Journal of \LaTeX\ Class Files,~Vol.~14, No.~8, August~2015}
{Dong \MakeLowercase{\textit{et al.}}: Towards Unified Multi-interest Recommendation through Prompt Learning}

\IEEEtitleabstractindextext{%
\begin{abstract}
Multi-interest learning method for sequential recommendation aims to predict the next item according to user multi-faceted interests given the user historical interactions. Existing methods mainly consist of a multi-interest extractor that embeds the multiple user interests based on the user interactions, and a multi-interest aggregator that aggregates the learned multi-interest embeddings to derive the final user embedding, used for predicting the user rating to an item. Despite their effectiveness, existing methods have two key limitations: 1) they directly feed the user interactions into the multi-interest extractor and aggregator, while ignoring their different learning objectives, and 2) they merely consider the centrality of the user interactions to embed multiple interests of the user, while overlooking their dispersion. To tackle these limitations, we propose a prompt-based multi-interest learning method (PoMRec), where specific prompts are inserted into user interactions, making them adaptive to the extractor and aggregator. Moreover, we utilize both the mean and variance embeddings of user interactions to embed the user multiple interests for the comprehensively user interest learning. We conduct extensive experiments on three public datasets, and the results verify that our proposed PoMRec outperforms the state-of-the-art multi-interest learning methods.
\end{abstract}

}

\maketitle

\IEEEdisplaynontitleabstractindextext

\IEEEpeerreviewmaketitle

\section{Introduction}
\IEEEPARstart{R}{ecommender} systems have become increasingly prevalent in real-world applications, which target on recommending items for users based on their interests. 
The core of the recommender systems is to learn the user and item embeddings, and predict the user rating to the item with the distance between their embeddings. 
Traditional recommendation methods~\cite{WuQLWZL19, XueHWXLH19, 0001DWLZ020} treat the user historical interactions as a set, ignoring their sequential information.
Intuitively, a user may purchase a phone case after buying a cellphone. Therefore, many researches\cite{HidasiKBT15, LiuWWLW16, KangM18, WangMZCLM21} have formalized the sequential recommendation task, where user interactions are treated as an ordered sequence to predict the next item that the user will be interested in.
Currently, the mainstream sequential recommendation methods focus on devising various neural networks, such as the recurrent neural networks~\cite{HidasiKBT15, LiuWWLW16} and Transformer~\cite{ZhaoZZL22}, to encode the user interaction sequence into a single embedding to represent the user interests.

However, user interests are diverse and \mbox{multi-faceted}. For example, a girl may be simultaneously interested in jewelry, handbags, and make-ups.
In such case, a single interest embedding could be hard to accurately capture all the user's diverse interests. 
Hence, several multi-interest learning methods~\cite{LiLWXZHKCLL19, CenZZZYT20, WangWLGM00FDM22, LiZBCSDJL22} for the sequential recommendation have been proposed to learn multiple interest embeddings for each user.  
Generally, these methods involve two key modules. 1) The \textit{multi-interest extractor} that aims to derive multiple interest embeddings for one userto capture his/her multi-faceted interests.
And 2) the \textit{multi-interest aggregator} that aggregates the learned multi-interest
embeddings to one user embedding. Then the user rating to one item can be predicted by the dot product between the user and item embeddings.
Despite their effectiveness, existing multi-interest learning methods for the sequential recommendation still have the following~limitations: 
\begin{itemize}[leftmargin=5mm]
\item \textit{Fail to adapt the user interactions to different learning objectives of the multi-interest extractor and aggregator.}
Since both the user multi-interest embeddings and their aggregation weights can be referred to the user interactions, existing methods directly take the user interactions as the inputs to the multi-interest extractor and aggregator.
In fact, the extractor focuses on the contents of user interactions to embed the user multiple interests, while the aggregator emphasizes analyzing the distribution of user interactions over multiple interests to derive the aggregation weights. Therefore, existing methods fail to make the inputted user interactions adaptive to different learning objectives of the two modules, whereby may derive the precise results. 
\item \textit{Overlook the dispersion of user interactions during the user multi-interest learning.}
Existing approaches mainly derive the user multi-interest embeddings by selecting a representative embedding of the user interaction embeddings, e.g., the weighted summation.
his representative embedding only capture the centrality of user interactions. However, the user interactions can be dispersed and it is insufficient that existing methods learn the user multi-interest embeddings with only their centrality when the user interactions become more dispersed. 
\end{itemize}

\begin{figure*}[!t]
\centering
\setlength{\abovecaptionskip}{0.2cm}
\includegraphics[width=\linewidth]{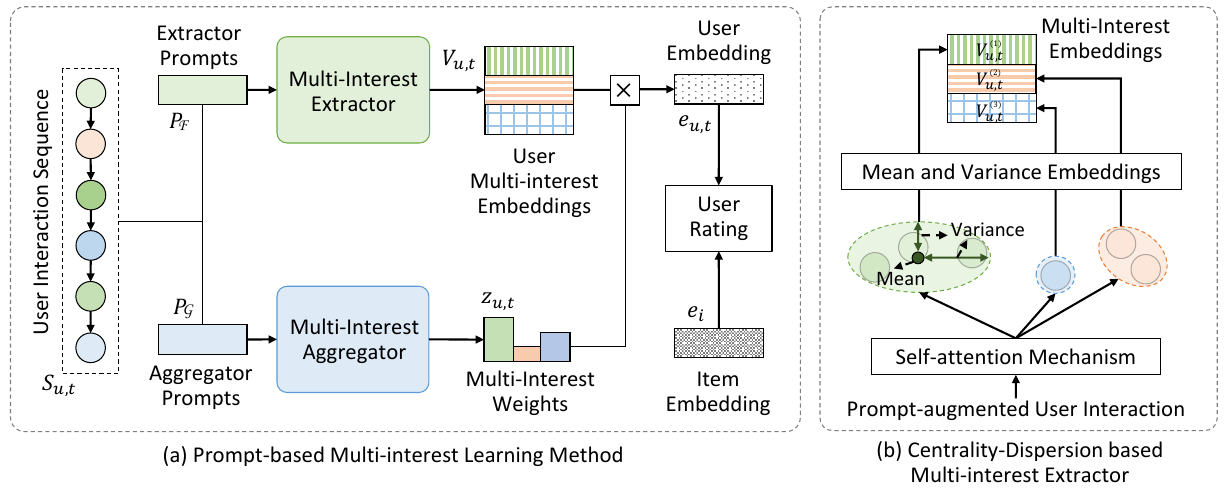}
\caption{Overview of the proposed prompt-based multi-interest learning method (PoMRec) for sequential recommendation (a). We insert certain prompt embeddings at the beginning of the user interaction sequence to make the model know whether it should focus on the contents of user interactions to
derived the multi-interest embeddings or the preference over multiple interest to predict the aggregation weights. Besides, we propose a centrality-dispersion based multi-interest extractor (b) that derives the multi-interests embeddings based on both
the centrality and dispersion of user interactions. Here we provide an example of the user that has three interests. }
\label{fig_method}
\end{figure*} 

To address the aforementioned limitations, in this paper, we propose a \textbf{P}r\textbf{o}mpt-based \textbf{M}ulti-interest learning method for the sequential \textbf{Rec}ommendation, termed as PoMRec, as shown in Figure~\ref{fig_method}.
As same as existing multi-interest learning methods, PoMRec consists a multi-interest extractor that learns the user multi-interest embeddings, and a multi-interest aggregator that learns the weights to fuse the multi-interest embeddings. 
Differently, in order to adapt the inputted user interactions to different learning objectives of the extractor and aggregator, we take inspiration from the soft prompts introduced in the prompt tuning~\cite{LesterAC21, LiuYFJHN23} that can make the original inputs adaptive to different downstream tasks. 
As can be seen in Figure~\ref{fig_method}~(a), we first introduce multiple learnable prompt embeddings for the multi-interest extractor and aggregator, respectively. 
Then the original inputs, i.e., the user interactions, augmented by the prompt embeddings, are fed into the two modules.
With the help of the prompts, the model will know whether it should focus on the contents of user interactions to derived the multi-interest embeddings or their distribution over multiple interests to derive the
aggregation weights. 
Thereafter, the outputted user multi-interest embeddings are aggregated by the weights as the final user embedding to predict the user rating to a given item. 
Besides, to embed the user multiple interests in a more comprehensive manner, we propose a centrality-dispersion based multi-interest extractor, as shown in Figure~\ref{fig_method}~(b), which attempts to learn the multi-interest embeddings considering both the centrality and dispersion of the user interactions. We first adopt the self-attention mechanism to softly cluster the prompt-augmented user interactions into several groups. Then we calculate both the mean and variance embeddings of the interaction embeddings in each group, and incorporate them as the final user interest embedding. 
We conduct extensive experiments on three public datasets: ML-1M, Beauty, and Movie \& TV. The experimental results have demonstrated the effectiveness of the proposed PoMRec. 

The main contributions can be summarized as follows:
\begin{itemize}[leftmargin=5mm]
\item We propose a prompt-based multi-interest learning method for the sequential recommendation (PoMRec) that introduces the learnable prompt embeddings into the inputted user interactions to make them adaptive to the different learning objectives of the multi-interest extractor and aggregator.
\item Different from existing methods that only consider the centrality of user interactions when embedding the user multi-interests, we propose a centrality-dispersion based multi-interest extractor that further takes the dispersion of user interactions as a supplement for seeking a better representation of the user multiple interests.
\item Extensive experiments on the three public datasets have demonstrated the effectiveness of the proposed method. We have released our codes to facilitate other researchers in https://github.com/hello-dx/PoMRec.
\end{itemize}
	
\section{Related work}
\label{section_related_work}
In this section, we first introduce the single-interest and multi-interest learning methods in Subsection~\ref{section_single_interest} and Subsection~\ref{section_multi_interest}, respectively. Besides, we review the prompt tuning methods in Subsection~\ref{section_prompt_tuning}.

\subsection{Single-interest Learning Method} 
\label{section_single_interest}
The sequential recommendation aims to predict the next item that the user might be interested in based on the user historical interaction sequence~\cite{HidasiKBT15, KangM18, SunLWPLOJ19, LiuFWY21}. Early researches~\cite{ChengYLK13, HidasiKBT15, KangM18} focus on encoding the user interaction sequence into a single embedding to represent the user single-interest. For example, Cheng~et~al.~\cite{ChengYLK13} adopted Markov chains to capture the correlation among items. With the development of deep neural networks, several studies employ the sequential modeling techniques, e.g., the Gate Recurrent Unit~\cite{HidasiKBT15} and Long Short-Term Memory~\cite{ZhouHHZT18}, to encode the user interaction sequence into a single embedding to predict the next item. Recently, the attention mechanism~\cite{KangM18, VaswaniSPUJGKP17, ShaW17} has shown promising potential to capture context-aware interests. However, user interests are diverse and multi-faceted. The aforementioned methods that utilize a single embeddings is hard to capture all the user's diverse interests. Besides, existing methods only leverage one embedding to represent the centrality of user interactions, which might be insufficient to capture the user interests when the user interactions are dispersed.

\subsection{Multi-interest Learning Method} 
\label{section_multi_interest}
As user interests are diverse, multi-interest learning has been proposed for the sequential recommendation, which aims to learn multiple embeddings for one user to capture the user multi-interests~\cite{MaZYCHC18, PiBZZG19, LiLWXZHKCLL19, CenZZZYT20, LiSZYZJYY23}. 
Existing methods generally involve a multi-interest extractor to learn the multiple interests of the user, and a multi-interest aggregator to fuse the multiple interests for generating the final recommendation.
In particular, as for the extractor, MIND~\cite{LiLWXZHKCLL19} uses the dynamic capsule routing method to group user interactions and obtain multiple user embeddings. Cen~et~al.~\cite{CenZZZYT20} adopted the self-attention mechanism to generate multiple embeddings from the inputted user interactions. 
As for the aggregator, several approaches utilize the simple greedy inference strategy~\cite{CenZZZYT20, ChenZZXX21} that utilizes the best matching interest
embedding (among all interest embeddings) to rank the item. Other recent approaches~\cite{MaZYCHC18, WangWLGM00FDM22} predict a weight vector to aggregate the user multi-interest embeddings into the user embedding, which shows greater performance. For example, Wang~et~al.~\cite{WangWLGM00FDM22} utilized  the Gate Recurrent Unit (GRU) to encode user interactions into one embedding and mapped it to the weight vector by the multi-layer perceptron.

Nevertheless, these methods directly feed the user interactions into the multi-interest extractor and aggregator, without making the inputs adaptive to the different learning objectives. In this paper, we introduce specific prompts into user interactions in order to make the downstream model known whether
it should focus on the contents of user interactions or the user preference
over multiple interests. 

\subsection{Prompt Tuning}
\label{section_prompt_tuning}
The prompt learning is firstly proposed to overcome the gap between the pre-training and fine-tuning~\cite{GuHLH22, LiuYFJHN23}. It focuses on adding prompts to the downstream tasks of the pre-training models to improve the performance of the downstream tasks without the fine-tuning step. Early approaches~\cite{BrownMRSKDNSSAA20, RaffelSRLNMZLL20} mostly incorporate manually generated discrete prompts to guide the model. Later, since the manually generated prompts are both time-consuming and trivial, other researches~\cite{GaoFC20,JiangXAN20, ShinRLWS20} turn to automatically search discrete prompts for specific tasks. Nevertheless, these methods largely depend on the quality of the generated prompts. Some recent approaches have begun to utilize the continuous learnable embeddings as the prompts~\cite{LesterAC21, LiangHXMHCGLJ21, GuoLY22} achieving the state-of-the-art performance. 

Inspired by them, in this paper, we introduce the prompt embeddings into the multi-interest interest learning to make the inputs adaptive to the downstream multi-interest extractor and aggregator. 
Our approach is the first attempt to add the prompt embeddings into the multi-interest learning for the sequential recommendation.

\section{Methodology}
\label{section_methodology}
We propose a prompt-based multi-interest learning method for sequential recommendation (PoMRec). 
In this section, we first brief the problem definition of the sequential recommendation in Subsection~\ref{section_problem_definition}. We then detail the proposed PoMRec in Subsection~\ref{section_method}, followed by the model complexity analysis in Subsection~\ref{section_complexity}.

\begin{table}[t]
\centering
\renewcommand{\arraystretch}{1.3}
\caption{Summary of the Main Notations.} 
\label{table_notations}
\setlength{\tabcolsep}{1.2mm}{
\begin{tabular}{c||l}	
    \hline
    Notation & Explanation \\
    \hline\hline
    $\mathcal{U}$, $\mathcal{I}$ & Sets of users and items, respectively. \\\hline
    $S_u$ & User~$u$'s interaction sequence. \\\hline
    $S_{u,t}$ & User~$u$'s interaction sequence truncated at time~$t$. \\\hline
    $K$ & Predefined number of user interests.\\\hline
    $d$ & Embedding size.\\\hline
    $\bm{e}_i \in \mathbb{R}^d$ & Embedding of the item~$i$.\\\hline
    $\bm{e}_{u,t} \in \mathbb{R}^{d}$ & Embedding of the user~$u$ at time~$t$. \\\hline
    \multirow{2}{*}{$\bm{H}_{u,t} \in \mathbb{R}^{d \times M}$} & Interaction embeddings~$\bm{H}_{u,t} = [\bm{e}_{u,t-M},...,\bm{e}_{u,t}]$ \\
    & of user~$u$ contains most recent $M$ items in $S_{u,t}$. \\\hline
    $N_p$& Number of the prompt embeddings. \\\hline
    $\bm{P}_\mathcal{F} \in \mathbb{R}^{N_p \times d}$ & Prompts for multi-interest extractor. \\\hline
    $\bm{P}_\mathcal{G}  \in \mathbb{R}^{N_p \times d}$ & Prompts for multi-interest aggregator. \\\hline
    \multirow{2}{*}{$\bm{V}_{u, t} \in \mathbb{R}^{d \times K}$} & Multi-interest embeddings of the user~$u$, where the  \\
     & column vector $\bm{V}_u^{(k)}$ is the $k$-th interest embedding. \\\hline
    $\bm{z}_{u,t} \in \mathbb{R}^{K}$ & Interest weights of the user~$u$ at time~$t$. \\\hline
    $y_{u,t}^i $ & Predicted rating of the user~$u$ to item~$i$ at time~$t$. \\\hline
    \multirow{2}{*}{$\lambda$} & Trade-off parameter between the centrality and dis- \\
     & persion of user interactions. \\\hline
\end{tabular}}	
\end{table}

\subsection{Problem Definition}
\label{section_problem_definition}
To improve the readability, we declare the notations used in this paper. We use the squiggled letters (\textit{e.g.,} $\mathcal{X}$) to represent sets. The bold capital letters (\textit{e.g.,} $\bm{X}$) and bold lowercase letters (\textit{e.g.,} $\bm{x}$) represent matrices and vectors, respectively. Let the nonbold letters (\textit{e.g.,} $x$) denote scalars. 

Suppose that there is a set of users~$\mathcal{U}$, and a set of items~$\mathcal{I}$. Each user~$u \in \mathcal{U}$ is associated with a sequence of all his/her historical interactions sorted by their corresponding interacted timestamps, denoted as~$S_u = [i_{u,1}, i_{u,2}, ..., i_{u,N_u}]$, where~$i_{u,t}$ is the interacted item at time step~$t$ and $N_u$ is the length of the list. Different from traditional recommendation that represents the user with a given user ID, the sequential recommendation resort to a sequence of item IDs that the user historically interacted.
In particular, as for a given user, the goal of the sequential recommendation takes the user interaction sequence~$S_{u,t}$ truncated as the time step~$t$ as the input and predicts the next item~$i_{u,t+1}$ that the user will be interested in be. The notations used in this paper are summarized in Table~\ref{table_notations}.

\subsection{Prompt-based Multi-interest Learning Method for the Sequential Recommendation (PoMRec)}
\label{section_method}
In this subsection, we first outline the proposed PoMRec method. Then we provide an implementation of the multi-interest extractor and aggregator in PoMRec, respectively, followed by the model optimization. 

\subsubsection{Overall Framework}
\label{section_overal_framework}
Given the user interaction sequence~$S_{u,t}$, PoMRec has a multi-interest extractor that derives the user multi-interest embeddings and a multi-interest aggregator that learns the weights to fuse the multi-interest embeddings. Different from previous methods, PoMRec inserts certain prompts into the inputted user interaction sequence~$S_{u,t}$ before feeding it into the multi-interest extractor and aggregator, which makes it adaptive to different learning objectives of the two modules. 
 
\textbf{Inputs.} Accordingly, the inputs of PoMRec consist of two parts: the original user interactions and newly introduced prompts for the multi-interest extractor and aggregator. 
\begin{itemize}[leftmargin=5mm]
\item To represent the original user interactions, following most recommendation methods~\cite{0001DWLZ020, CenZZZYT20, WangWLGM00FDM22}, we represent each item~$i \in \mathcal{I}$ with a $d$-dimensional learnable embedding~$\bm{e}_i \in \mathbb{R}^d$. Then the user historical interaction sequence~$S_{u,t}$ can be transformed into an embedding sequence~$\bm{H}_{u,t} = [\bm{e}_{u,t-M+1},...,\bm{e}_{u,t-1},\bm{e}_{u,t}]$, which consists of the most recent interacted $M$ items at the time step~$t$. We term the embedding sequence~$\bm{H}_{u,t}$ as the user interaction embeddings. 
Notably, if the sequence length is less than $M$, we repeatedly add a zero embedding to the beginning of the sequence until the length becomes $M$. In addition, we also add the trainable positional embeddings to each item embedding in~$\bm{H}_{u,t}$ in order to make use of the sequential information. 
\item We then introduce the prompts for the multi-interest extractor and aggregator. Following the soft prompts in prompt tuning methods~\cite{LesterAC21, LiangHXMHCGLJ21}, we introduce the to-be-learned prompt embeddings $\bm{P}_\mathcal{F}= [\bm{p}_\mathcal{F}^{1},..., \bm{p}_\mathcal{F}^{N_p}]$ for the extractor, and $\bm{P}_\mathcal{G} = [\bm{p}_\mathcal{G}^{1},..., \bm{p}_\mathcal{G}^{N_p}]$ for the aggregator, both of which consist of $N_p$ randomly initialized \mbox{$d$-dimensional} embeddings. The prompt embeddings $\bm{P}_\mathcal{F}$ and $\bm{P}_\mathcal{G}$ are expected to capture the specific learning objectives of the extractor and aggregator, respectively. 
Notably, we use more than one prompt embedding to increase the adaptive ability of the inputs~\cite{WangZWZ22}. 
\end{itemize}

Ultimately, the prompt embeddings can be as the identifier to prompt the model to determine whether it should focus on the contents or the distribution over interests of user interactions. 
Formally, we insert the prompt embeddings at the beginning of the interaction embeddings~$\bm{H}_{u, t}$ as the inputs for the extractor and aggregator as follows,
\begin{equation}
\label{eqn_prompt_embeddings}
\begin{cases}
\bm{H}^\mathcal{F}_{u, t} = [\bm{P}_\mathcal{F}, \bm{H}_{u, t}], \\
\bm{H}^\mathcal{G}_{u, t} = [\bm{P}_\mathcal{G}, \bm{H}_{u, t}],
\end{cases}
\end{equation}
where $\bm{H}^\mathcal{F}_{u, t} \in \mathbb{R}^{(N_p+M)\times d}$ and $\bm{H}^\mathcal{G}_{u, t} \in \mathbb{R}^{(N_p+M)\times d}$ are the prompt-augmented interaction embeddings inputted into the multi-interest extractor and aggregator, respectively. 

\textbf{Centrality-Dispersion based Multi-interest Extractor.}
Considering that the user's current interests can be inferred by his/her recent interactions, this extractor~$\mathcal{F}$ will output the user multi-interest embeddings based on the prompt-augmented interaction embeddings~$\bm{H}^\mathcal{F}_{u, t}$. 
Formally, suppose that each user has $K$ interests. The user multi-interest embeddings~$\bm{V}_{u, t}$ can be derived as follows,
\begin{equation}
\label{eqn_interest_embeddings}
\bm{V}_{u, t} = \mathcal{F} (\bm{H}^\mathcal{F}_{u, t}|\bm{\Theta}_\mathcal{F}),
\end{equation}
where $\bm{V}_{u, t} \in \mathbb{R}^{d\times K}$ refer to the total $K$ interest embeddings of the user~$u$ at the time step~$t$. $\bm{\Theta}_\mathcal{F}$ is the set of parameters in the extractor~$\mathcal{F}$. The detailed implementation of the extractor can refer to Subsection~\ref{section_multiinterest_extractor}.

\textbf{Attention-based Multi-interest Aggregator.} The aggregator~$\mathcal{G}$ focuses on aggregating the learned multi-interest embeddings~$\bm{V}_{u, t}$ to one embedding for predicting the final user rating. In particular, the aggregation weights of different user interests can be traced from the user interactions. For example, if the user have purchased the outdoor equipment, he/she is not likely to prefer the same items in the next time. 
Therefore, we feed the prompt-augmented interaction embeddings~$\bm{H}^\mathcal{G}_{u, t}$ into the aggregator~$\mathcal{G}$ to learn the weights of multiple user interests. Then the final user embedding $\bm{e}_{u, t} \in \mathbb{R}^d$ at the time step~$t$ can be aggregated by the learned weights as follows,
\begin{equation}
\label{eqn_aggregation}
\bm{e}_{u, t} = \mathcal{G} (\bm{H}^\mathcal{G}_{u, t}, \bm{V}_{u, t}|\bm{\Theta}_\mathcal{G}),
\end{equation}
where $\bm{\Theta}_\mathcal{G}$ is the set of parameters in the aggregator~$\mathcal{G}$. The detailed implementation can refer to Subsection~\ref{section_multiinterest_aggregator}.

\textbf{Output}. Ultimately, we use the commonly-used dot product between the user and item embeddings as the rating of the user~$u$ to a given item~$i$ as follows, 
\begin{equation}
\label{eqn_user_rating}
y_{u, t}^i = \bm{e}_{u, t}^\mathsf{T} \bm{e}_i,
\end{equation}
where $y_{u, t}^i$ is the rating of the user~$u$ to the item~$i$ at the time step~$t$. 
Accordingly, based on the predicted user ratings, the candidate items can be ranked and top items are recommended to the user. 

\subsubsection{Centrality-Dispersion Based Multi-interest Extractor}
\label{section_multiinterest_extractor}
The basic idea of the centrality-dispersion based multi-interest extractor is to first represent the centrality of each interest in the user interactions and then calculate the dispersion according to the centrality as the supplement.
To be more specific, following studies~\cite{CenZZZYT20, WangWLGM00FDM22}, we adopt the self-attention mechanism to learn the centrality representation matrix~$\bm{M}_{u, t} \in \mathbb{R}^{d\times K}$ of the user multi-interests as follows,
\begin{equation}
\label{eqn_new_interest_embedding}
\begin{cases}
 \bm{M}_{u, t} = (\bm{H}^\mathcal{F}_{u, t})^\mathsf{T} \bm{A}^{\mathcal{F}}_{u, t},\\
 \bm{A}^\mathcal{F}_{u, t} = {\rm softmax} \Big(\bm{W}_\mathcal{F}^2 {\rm tanh}(\bm{W}_\mathcal{F}^1  \bm{H}^\mathcal{F}_{u, t})\Big)^\mathsf{T},\\
\end{cases}
\end{equation}
where 
$\bm{A}^\mathcal{F}_{u, t} \in \mathbb{R}^{(N_p+M)\times K}$ is an attention matrix, which represents the probability of each user interaction belonging to each interest of the user~$u$. Intuitively, the attention matrix is able to softly classify the user interactions into $K$ groups. Accordingly, the $k$-th column of the centrality representation matrix, i.e, $\bm{M}_{u, t}^{(k)}$ is derived by the weighted summation of the interaction embeddings, which represents the centrality of the $k$-th interests of the user~$u$.
The function ${\rm softmax}$ is to enforce the summation of the elements in the $k$-th column vector of the attention matrix to equal to $1$.
$\bm{W}_\mathcal{F}^1  \in \mathbb{R}^{d' \times d}$ and $\bm{W}_\mathcal{F}^2 \in \mathbb{R}^{K \times d'}$ are the trainable parameters. ${\rm tanh}$ is the activation function. It is worth noting that we can also utilize other approaches to learn the centrality representation matrix~$\bm{M}_{u, t}$.

\begin{figure}[!t]
\centering
\includegraphics[width=\linewidth]{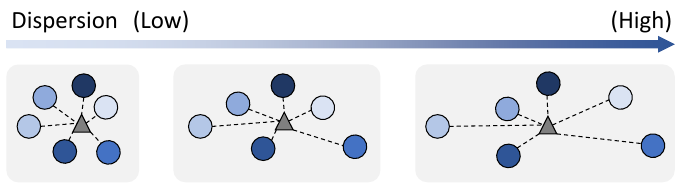}
\caption{Illustration of the user interactions with different dispersion. The blue points represent the user interacted items, and the grey triangles are their corresponding centrality representation. Intuitively, along with the dispersion of user interactions from low to high, the reliability of the centrality for representing the user interactions decreases.}
\label{fig_dispersion}
\end{figure} 

Existing approaches directly utilize the centrality representation matrix~$\bm{M}_{u, t}$ as the user multi-interest embeddings. 
Nevertheless, as shown in Figure~\ref{fig_dispersion}, along with the user interactions becoming increasingly dispersed, the reliability of the centrality representation to represent the user interactions tends to decrease. To tackle this issue, in this work, we first calculate a dispersion representation matrix based on the centrality representation matrix, and then derive the use multi-interest embeddings by their combination. 
Technically, we define the multi-interest embeddings~$\bm{V}_{u, t} \in \mathbb{R}^{d\times K}$ of the user~$u$ as follows,
\begin{equation}
\label{eqn_new_interest_embedding}
\begin{cases}
\bm{V}_{u, t} = \bm{M}_{u, t} + \lambda \bm{\Sigma}_{u, t}, \\
 \bm{\Sigma}_{u, t}^{(k)} = {\rm sqrt}\Big(((\bm{H}^\mathcal{F}_{u, t})^2)^\mathsf{T} \bm{A}^{\mathcal{F}(k)}_{u, t} - (\bm{M}_{u, t}^{(k)})^2\Big), k=1,...,K,\\
\end{cases}\
\end{equation}
where $\bm{\Sigma}_{u, t}  \in \mathbb{R}^{d\times K}$ is the dispersion representation matrix, calculated by the the dispersion of user interactions from each centrality. 
$\lambda$ is a hyper-parameter to adjust the trade-off between the centrality and dispersion of the user interactions in learning the user multi-interests.
The superscript~$^2$ of a matrix denotes the element-wise multiplication of the matrix.  $\bm{M}_{u, t}^{(k)}$ is the $k$-th column vector of the matrix~$\bm{M}_{u, t}$. 

\subsubsection{Attention-based Multi-interest Aggregator}
\label{section_multiinterest_aggregator}
The aggregator~$\mathcal{G}$ first predicts a weight vector that captures the weights of user multiple interests, and then aggregates the user multi-interest embeddings~$\bm{V}_{u, t}$ to the final user embedding. 
To fulfill this, we first aggregate these embeddings into a single embedding, and then map the embedding into a \mbox{$K$-dimensional} vector as the weight vector. 

Technically, due to the impressive ability of self-attention mechanism in learning the attention weight of the embedding aggregation, we adopt it to aggregate the prompt-augmented interaction embeddings~$\bm{H}^\mathcal{F}_{u, t}$ to derive a summarized embedding~$\bm{v}_{u, t} \in \mathbb{R}^d$ as follows, 
\begin{equation}
\begin{cases}
\bm{a}^\mathcal{G}_{u, t} = {\rm softmax} \Big(\bm{W}^2_\mathcal{G} {\rm tanh}(\bm{W}^1_\mathcal{G} \bm{H}^\mathcal{G}_{u, t})\Big)^\mathsf{T},\\
\bm{v}_{u, t} = \bm{H}^\mathcal{G}_{u, t} \bm{a}^\mathcal{G}_{u, t},
\end{cases}
\end{equation}
where $\bm{a}^\mathcal{G}_{u, t} \in \mathbb{R}^{N_p+M}$ contains the attention weights of the prompt-augmented interaction embeddings to derive the summarized embedding.
$\bm{W}_\mathcal{G}^1  \in \mathbb{R}^{d' \times d}$ and $\bm{W}_\mathcal{G}^2 \in \mathbb{R}^{1 \times d'}$ are the trainable parameters. 

We then adopt the multi-layer perceptron to map the summarized embedding to the weight vector as follows,
\begin{equation}
\bm{z}_{u, t} = \underbrace{ \cdots \bm{W}^2_M {\rm tanh}(\bm{W}^1_M}_L \bm{v}_{u, t} + \bm{b}^1_M) + \bm{b}^2_M \cdots,
\end{equation}
where~$\bm{z}_{u, t} \in \mathbb{R}^K$ denotes the weights of multiple interests of the user~$u$ at the time step~$t$. $L$ is the layer of the multi-layer perceptron, which is set to $2$ empirically. $\bm{W}^1_M \in \mathbb{R}^{d'\times d}$, $\bm{b}^1_M \in \mathbb{R}^{d'}$, $\bm{W}^2_M \in \mathbb{R}^{d'\times K}$ and $\bm{b}^2_M \in \mathbb{R}^{K}$ are the trainable parameters.

Accordingly, we aggregate the learned user multi-interest embeddings $\bm{V}_{u, t}$ based on the weight vector~$\bm{z}_{u, t}$ to derive the user embedding as follows,
\begin{equation}
\bm{e}_{u, t} =  \bm{V}_{u, t} \bm{z}_{u, t},
\end{equation}
where $\bm{e}_{u, t} \in \mathbb{R}^{d}$ captures the user multi-interests at the time step~$t$, which is used to predict the final user rating for the recommendation.

\subsubsection{Model Optimization}
\label{section_optimization}
The goal of the sequential recommendation is to predict the next item given the user interaction sequence at one time step. 
Accordingly, following Bayesian personalized ranking mechanism~\cite{RendleFGS09}, we build the following training~set,
\begin{equation}
\label{eqn_training_set}
\mathcal{D} = \Big\{ (S_{u, t},i,j) \;\;
\begin{array}{|l}
 u \in\mathcal{U}, t = 1, ..., N_u-1, \\
 i=i_{u, t+1} \in S_u, j\in \mathcal{I}\setminus S_u,
\end{array}
\Big\}
\end{equation}
where $S_{u,t}$ is the user interaction sequence truncated at time step~$t$, $i= i_{u, t+1}$ is the target next item in~$S_u$, and $j\in \mathcal{I}\setminus S_u$ is a negative item that randomly sampled from items that the user has not interacted. The training triplet~$(S_{u, t}, i, j)$ indicates that the user~$u$ prefers item~$i$ compared to the item~$j$ at the time step~$t$. We then adopt the pair-wise objective function to ensure the rating of the user~$u$ to the positive item~$i$ is larger than the negative item~$j$ as follows,
\begin{equation}
\label{eqn_mf_loss}
    \mathcal{L} = \min_{\bm{\Theta}} -\sum_{\mathcal{D}} {\rm log}\Big({\rm sigmoid}(y_{u,t}^i - y_{u,t}^j)\Big),
\end{equation}
where $y_{u,t}^i$ and $y_{u,t}^j$ are the ratings of the user~$u$ to the item~$i$ and~$j$  at the time step~$t$, which can be derived in Eqn~(\ref{eqn_user_rating}). ${\rm sigmoid}$ is the sigmoid activation function. $\bm{\Theta}$ is the set of all parameters in the method, consisting of the item embeddings~$\{\bm{e}_i\}_{i \in \mathcal{I}}$, the prompt embeddings~$\bm{P}_\mathcal{F}$ and $\bm{P}_\mathcal{G}$, and parameters in the multi-interest extractor~$\bm{\Theta}_{\mathcal{F}}$ and target-interest predictor~$\bm{\Theta}_{\mathcal{G}}$.
The detailed training process of the proposed PoMRec is summarized in Algorithm~\ref{algorithm_training}.

\begin{algorithm}[t]
\caption{Training Process of the Proposed PoMRec.}
\label{algorithm_training}
\begin{algorithmic}[1]
    \REQUIRE 
    The set of users~$\mathcal{U}$ and set of items~$\mathcal{I}$.
    The historical interactions~$S_u$ of each~$u \in \mathcal{U}$. The training set~$\mathcal{D}$. \\
    The hyper-parameters~$K$, $N_P$, and $\lambda$. 
    \ENSURE The model parameters~$\boldsymbol{\Theta}$.\\
    \STATE Randomly initialize the embeddings~$\{\bm{e}_i\}_{i \in \mathcal{I}}, \bm{P}_\mathcal{F}, \bm{P}_\mathcal{G}$ and parameters~$\bm{\Theta}_{\mathcal{F}}, \bm{\Theta}_{\mathcal{G}}$. 
    \STATE Shuffle the training triplets $(S_{u, t},i,j)$ in~$\mathcal{D}$.
    \WHILE {not converged}
    \STATE Draw a mini-batch of training triplet~$(S_{u, t},i,j)$.
    \STATE Construct the prompt-augmented user interaction embeddings $\bm{H}^\mathcal{F}_{u, t}$ and $\bm{H}^\mathcal{G}_{u, t}$.
    \STATE Calculate the user multi-interest embeddings: \\
    $\bm{V}_{u, t} = \mathcal{F} (\bm{H}^\mathcal{F}_{u, t}|\bm{\Theta}_\mathcal{F})$. \\
    \STATE Calculate the final user embedding: \\
    $\bm{e}_{u, t} = \mathcal{G} (\bm{H}^\mathcal{G}_{u, t}, \bm{V}_{u, t}|\bm{\Theta}_\mathcal{G})$. \\
    \STATE Calculate the user ratings to the item~$i$ and $j$: \\
    $y_{u, t}^i = \bm{e}_{u, t}^\mathsf{T} \bm{e}_i, y_{u, t}^j = \bm{e}_{u, t}^\mathsf{T} \bm{e}_j.$ \\
    \STATE Update the parameters of PoMRec: \\
    $\boldsymbol{\Theta} \leftarrow \boldsymbol{\Theta}-\eta \frac{\partial \mathcal{L}}{\partial\boldsymbol{\Theta}} $\\
    \ENDWHILE
    \RETURN Model parameters~$\boldsymbol{\Theta}$.
\end{algorithmic}
\end{algorithm}

\subsection{Model Complexity Analysis}
\label{section_complexity}
In PoMRec, the main time-consuming steps appear in the user multi-interest extraction and multi-interest weight prediction, i.e., step $6$ and step $7$ in Algorithm~$1$. 
\begin{itemize}[leftmargin=5mm]
    \item Step $6$ in Algorithm~$1$ learns the user multi-interest embeddings, and it has computational complexity $\mathcal{O}((dd'+d'K)(M+N_p))$ to derive the attention matrix~$\bm{A}_{\mathcal{F}u}$. Based on the attention matrix, we then calculate the mean and variance representations of the user multi-interests, both of which have the computational complexity $\mathcal{O}(dK(M+N_p))$. Accordingly, the user multi-interest embeddings can be calculated through Eqn.~(\ref{eqn_new_interest_embedding}) with the computational complexity $\mathcal{O}(d(M+N_p))$. 
    \item Step $7$ in Algorithm~$1$ learns the weight vector and it first derives the attention matrix~$\bm{A}_{\mathcal{G}u}$, which has the computational complexity~$\mathcal{O}(dd'(M+N_p))$. The weight vector then can be derived by the multi-layer perceptron taking the computational complexity $\mathcal{O}(dd'+d'K)$.
\end{itemize}

Therefore, the overall complexity for evaluating PoMRec is~$\mathcal{O}(d'dK(M+N_p))$.
Notably, compared with existing multi-interest learning methods, the proposed PoMRec only additionally introduces $N_p$ prompt embeddings for multi-interest extraction and multi-interest weight prediction, respectively, which burdens little to the space complexity.

\section{Experiments}
In this section, we first detail the experimental settings in Subsection~\ref{section_settings}, and then present the experiment results to answer the following research questions:

\begin{itemize}[leftmargin=9mm]
    \item[RQ1:] Does the proposed PoMRec outperform existing state-of-the-art methods?
    \item[RQ2:] How effective are the different proposals of PoMRec?
    \item[RQ3:] How effective is PoMRec deployed on existing multi-interest learning methods?
    \item[RQ4:] How do hyper-parameters affect model performance?
    \item[RQ5:] Do the learned multi-interest embeddings capture the user multi-interests?
\end{itemize}

\subsection{Experimental Settings}
\label{section_settings}
\textbf{Datasets.} To evaluate our proposed PoMRec in the task of the top-$N$ item recommendation, we conducted extensive experiments on the following three public datasets: \textbf{ML-1M}~\cite{HarperK16}, \textbf{Bueaty}~\cite{NiLM19} 
and \textbf{Movies \& TV}~\cite{NiLM19}, which have various scales and sparsities. All these datasets contain the users' ratings on items, and each rating is annotated with its time step. 
For the fair comparison, we closely followed the data pre-processing of the study~\cite{WangWLGM00FDM22}. 
The statistics of the datasets are summarized in Table~\ref{table_dataset}.

\textbf{Evaluation Protocols.} 
We adopt the leave-one-out evaluation to evaluate model performance, which is widely-used in previous studies~\cite{KangM18, WangWLGM00FDM22}. In particular, for each user in the dataset, the most recent interaction is used for testing, the second most recent interaction is used for validation, and the remaining interactions are for training. In the validation and testing, the candidate items consist of one ground-truth item and $999$ randomly sampled items that the user has not interacted with. This sampled metric is shown to be comparable to using the whole set of items as candidate items~\cite{LiJGL20, WangWLGM00FDM22}, while largely saving the computational time. 
We adopted Recall@$N$ and NDCG@$N$ to evaluate the effectiveness of our proposed PoMRec in the top-$N$ recommendation. By default, we set $N=5, 10, 20$.

\begin{table}[t]
\centering
\renewcommand{\arraystretch}{1.5}
\caption{Statistics of datasets.} 
\label{table_dataset}
\setlength{\tabcolsep}{2.5mm}{
    \begin{tabular}{c|c|c|c|c}	
        \hline
        \multirow{2}{*}{Dataset}& \#User & \#Item & \#Interaction & Density \\
         & $|\mathcal{U}|$ & $|\mathcal{I}|$ & $\sum_u N_u$ & $\frac{\sum_u N_u}{|\mathcal{U}|\times |\mathcal{I}|}$ \\ 
        \hline\hline   
        ML-1M & 6,040 & 3,706 & 1,000,209 & 4.47\% \\\hline
        Beauty & 22,363 & 12,101 & 198,502 & 0.07\% \\\hline
        Movies \& TV & 123,960 & 50,052 & 8,765,568 & 0.14\% \\\hline
\end{tabular}}	
\end{table}

\textbf{Implementation Configuration.} 
The hyper-parameters in our proposed method consist of the embedding size~$d$, the length~$M$ of the embedding sequence~$\bm{H}_{u,t}$, the number of user interests~$K$, the number of prompt embeddings~$N_p$, and the trade-off parameter~$\lambda$. 
In particular, we set the embedding size~$d$ to $64$ and length~$M$ to $20$ following existing studies~\cite{CenZZZYT20, ChenZZXX21, WangWLGM00FDM22}. 
We tuned these hyper-parameters from $\{1,2,3,4,5\}$, and results are discussed in Subsection~\ref{section_parameter}. 
We optimized the proposed PoMRec with Adam optimizer~\cite{KingmaB14}. We set the mini-batch sizes to $256$ for all the datasets. 
We trained the network with $200$ epochs with the early stop strategy, and selected the best model according to the performance of the validation set. Typically, $200$ epochs are sufficient for PoMRec to converge. 

\begin{table*}[]\small
\centering
\caption{Experimental results of the performance comparison. The best and second best results are in bold and underlined, respectively. The parameter numbers of the methods are listed in the ``Param.'' column. The superscript~$*$ denotes that the results of the method are referred to the study~\cite{WangWLGM00FDM22}.}
\label{table_comparison}
\renewcommand{\arraystretch}{1.3}
\setlength{\tabcolsep}{4.5mm}{
\begin{tabular}{crccccccc}
\hline
\multirow{2}{*}{Dataset} & \multirow{2}{*}{Method} &  \multicolumn{3}{c}{Recall@N} & \multicolumn{3}{c}{NDCG@N} & \multicolumn{1}{c}{Param.} \\ \cmidrule(lr){3-5}\cmidrule(lr){6-8}
& & N=5 & N=10 & N=20 & N=5 & N=10 & N=20 & (Million) \\ \hline
\multicolumn{1}{c}{\multirow{6}{*}{ML-1M}}&GRU4Rec$^*$ &  0.2730 & 0.3964 & 0.5323 & 0.1875 & 0.2273 & 0.2616 & 0.3 \\ 
&MIND$^*$    & 0.1863 & 0.2881 & 0.4152 & 0.1229 & 0.1558 & 0.1877 & 0.2  \\ 
&ComiRec$^*$ & 0.2513 & 0.3659 & 0.4937 & 0.1708 & 0.2078 & 0.2400 & 0.2 \\
&MINER  & 0.2758 & 0.3901 & 0.3901 & 0.1878 & 0.2246 & 0.2574 & 0.2 \\
&TiMiRec$^*$ & \underline{0.3091} & \underline{0.4310} & \underline{0.5625} & \underline{0.2136} & \underline{0.2529} & \underline{0.2861} & 0.5 \\
&PoMRec  & \textbf{0.3151} & \textbf{0.4422} & \textbf{0.5752} & \textbf{0.2188} &  \textbf{0.2598} & \textbf{0.2933} & 0.2 \\ \hline
\multicolumn{2}{r}{Relative Improvement} & 1.94\% & 2.60\% & 2.26\% & 2.43\% & 2.73\% & 2.52\% & -\\ \hline
\multicolumn{1}{c}{\multirow{6}{*}{Beauty}}&GRU4Rec$^*$ & 0.1072 & 0.1552 & 0.2107 & 0.0719 & 0.0873 & 0.1013 & 0.9 \\ 
&MIND$^*$ & 0.1193 & 0.1727 & 0.2492 & 0.0809 &  0.0981 & 0.1173 & 0.8 \\
&ComiRec$^*$ & 0.1257 & 0.1832 & 0.2543 & 0.0852 & 0.1038 & 0.1217 & 0.8 \\
&MINER & 0.1224 & 0.1740  & 0.2408 & 0.0841 & 0.1008 & 0.1176 & 0.8  \\
&TiMiRec$^*$  & \underline{0.1437} & \underline{0.2006} & \underline{0.2645} & \underline{0.1006} & \underline{0.1118} & \underline{0.1350} & 1.6 \\
&PoMRec   & \textbf{0.1456} & \textbf{0.2031} & \textbf{0.2713} & \textbf{0.1010} & \textbf{0.1195} & \textbf{0.1367} & 0.8 \\ \hline
\multicolumn{2}{r}{Relative Improvement} & 1.32\% & 1.25\% & 2.57\% & 0.40\% & 6.89\% & 1.26\% & -\\ \hline
\multicolumn{1}{c}{\multirow{6}{*}{Movies \& TV}} &GRU4Rec &  0.1928 & 0.2811 & 0.3871 & 0.1299 & 0.1584 & 0.1851 & 3.3 \\ 
&MIND    & 0.2394 & 0.3290 & 0.4282 & 0.1655 & 0.1954 & 0.2205 & 3.2   \\
&ComiRec & 0.2411 & 0.3291 & 0.4275 & 0.1687 & 0.1970 & 0.2219 & 3.2  \\
&MINER & 0.2467& 0.3398  & 0.4426 & 0.1702 & 0.2002 & 0.2262 & 3.2  \\
&TiMiRec  & \underline{0.2506} & \underline{0.3412} & \underline{0.4386} & \underline{0.1755} & \underline{0.2048} & \underline{0.2294} & 6.4 \\
&PoMRec   & \textbf{0.2747} & \textbf{0.3684} & \textbf{0.4714} & \textbf{0.1944} & \textbf{0.2246} & \textbf{0.2507} & 3.2 \\ \hline
\multicolumn{2}{r}{Relative Improvement} & 9.62\% & 7.97\% & 7.48\% & 10.77\% & 9.29\% & 8.67\% & -\\ \hline
\end{tabular}}
\end{table*}

\subsection{Performance Comparison (RQ1)}
\label{section_performance_comparison}
To quantitatively examine the effectiveness of the proposed PoMRec, we compared it with the following multi-interest learning methods for the sequential recommendation:
\begin{itemize}[leftmargin=5mm]
\item \textbf{GRU4Rec}~\cite{HidasiKBT15}is the first sequential recommendation method that utilizes GRU to model the user interactions into one user interest embedding.
\item \textbf{MIND}~\cite{LiLWXZHKCLL19} uses the dynamic routing mechanism in capsule networks to group the user interacted items into multiple clusters and obtain multiple user interest embeddings for recommendation. To generate the final recommendation, it utilizes the maximal matching score between the items and the user interests as the final rating of the user to the item, which is referred to the greedy inference strategy.
\item \textbf{ComiRec}~\cite{CenZZZYT20} uses multi-head attention mechanisms to learn multiple embeddings for each user to capture their diverse interests. It also utilizes the greedy inference strategy to generate the final recommendation. 
\item \textbf{MINER}~\cite{LiZBCSDJL22} introduces a novel poly attention
scheme to learn the user multi-interest embeddings. It also leverages a disagreement regularization to improve the poly attention, which enlarges the distance among different interest embeddings during training. The learned multi-interest embeddings are fed into the mean aggregation to derive the user embedding.
\item \textbf{TiMiRec}~\cite{WangWLGM00FDM22} applies the self-attention mechanism to the multi-interest extractor to learn the user's multi-interests and adopts the GRU to learn the weight vector from the user interactions. This method is a two-stage method, which first pre-trains the multi-interest extractor with the greedy inference strategy and then fine-tunes the interest weight predictor to learn the user embedding for predicting the recommended items.
\end{itemize}

Table~\ref{table_comparison} shows the comparison results of all baselines and the proposed PoMRec, where the best results are in bold and the second-best results are underlined. Besides, we also listed the total parameters of each method in Table~\ref{table_comparison}. From Table~\ref{table_comparison}, we have the following observations:
\begin{enumerate}[leftmargin=5mm]
\item Our proposed PoMRec achieves the best performance with respect to all metrics on both datasets. This demonstrates the effectiveness of our proposed method. The reasons behind this may be as follows: (i) by adding the specific prompts of learning objectives, the inputs of the user interactions are adaptive to the multi-interest extractor and aggregator, thus that the multi-interest embeddings and their weights can be better learned;
and (ii) we utilize both centrality and dispersion of the user interactions to learn the user multi-interest embeddings, which comprehensively models the user multi-interests.
\item The proposed PoMRec not only achieves the better performance but also needs significantly fewer parameters compared to the best baseline, i.e., TiMiRec. This is because that TiMiRec allocates two embeddings for each item, with one embedding used for the multi-interest extractor and another used for the multi-interest aggregator. In contrast, the proposed PoMRec only allocates one embedding for each item, and inserts specific prompt embeddings into the item embeddings in the user interaction sequence to make them adaptive to both the extractor and aggregator. 
\item Multi-interest learning methods MINER, TiMiRec, and TiMiRec perform better than the multi-interest learning methods MIND and ComiRec. This may be due to that although MIND and ComiRec learn the user multi-interests, they merely generate recommendations based on the best matching interest of the target item. This suggests that it is helpful to predict the next items that the user might be interested in by leveraging all the learned user interests. Moreover, MINER performs worse than TiMiRec and our proposed PoMRec. This may be because that MINER only utilize the mean aggregation of the learned user multi-interest embeddings to retrieve items, while TiMiRec and PoMRec utilize networks to predict interest weights to aggregate the learned user multi-interest embeddings.
\item Note that introducing the prompt embeddings could involve additional parameters, but can be negligible compared to the whole model parameters. Specifically, the model parameters largely depends on the number of items. There are 3,706, 12,101, and 50,052 items in ML-1M, Beauty and Movies \& TV datasets, respectively. However, there are at most 10 prompt embeddings introduced, i.e., there are two tasks in our method and we assign at most 5 embeddings for each task. Therefore, the new involved parameters only account for of $0.26\%$, $0.06\%$, and $0.02\%$ of original model parameters in ML-1M, Beauty and Movies \& TV datasets, respectively.
\end{enumerate}

\begin{table}[]\small
\caption{Experimental results of the ablation study on PoMRec. ``+Prompt'' denotes that we use the prompt embeddings to augment the inputted user interactions. ``+Disp.'' denotes that we utilize both the centrality and dispersion of user interaction to learn the multi-interest embeddings.}
\centering
\label{table_ablation_study}
\renewcommand{\arraystretch}{1.3}
\setlength{\tabcolsep}{2.5mm}{
\begin{tabular}{crcccc}
\hline
\multirow{2}{*}{Dataset} & \multirow{2}{*}{Variant} & \multicolumn{2}{c}{Recall@N} & \multicolumn{2}{c}{NDCG@N} \\ \cmidrule(lr){3-4} \cmidrule(lr){5-6}
 & & N=5 & N=10 & N=5 & N=10 \\ \hline
 \multirow{4}{*}{ML-1M} 
& Base & 0.3068 & 0.4248 & 0.2101 & 0.2484 \\  
& +Prompt & 0.3129 & 0.4333 & 0.2148 & 0.2536\\ 
& +Disp. & 0.3108 & 0.4320 & 0.2151 & 0.2544 \\
& PoMRec & \textbf{0.3151} & \textbf{0.4422} & \textbf{0.2188} &  \textbf{0.2598} \\  \hline
\multirow{4}{*}{Beauty}
& Base &  0.1193 & 0.1726 &  0.0823 & 0.0996 \\ 
& +Prompt &  0.1299 & 0.1838 & 0.0888 & 0.1061 \\
& +Disp. & 0.1293 & 0.1811 & 0.0889 & 0.1057 \\
& PoMRec & \textbf{0.1456} & \textbf{0.2031} & \textbf{0.1010} & \textbf{0.1195} \\ \hline
& Base &  0.2429 & 0.3295 & 0.1704 & 0.1983 \\ 
Movies& +Prompt &  0.2498 & 0.3406 & 0.1738 & 0.2031\\
\& TV& +Dis. & 0.2591 & 0.3536 & 0.1801 & 0.2106 \\
& PoMRec & \textbf{0.2747} & \textbf{0.3684} & \textbf{0.1944} & \textbf{0.2246} \\ \hline
\end{tabular}}
\end{table}

\subsection{Ablation Study (RQ2)}
In this subsection, we evaluated the proposed centrality-dispersion based multi-interest extractor and the prompt-based multi-interest learning method. 
In particular, we designed the following variants of PoMRec:
\begin{itemize}[leftmargin=5mm]
\item \textbf{Base.} We removed both the CD and BP in the proposed PoMRec. Specifically, we directly fed the user interaction embeddings~$\bm{H}_{u, t}$ without the prompts embeddings to both the multi-interest extractor and aggregator. Besides, in the multi-interest extractor, we only utilized the centrality representation as the user multi-interest embeddings, i.e.,~$\bm{V}_{u, t} = \bm{M}_{u,t}$, without the dispersion representation~$\bm{\Sigma}_{u,t}$.
\item \textbf{+Prompt.} This variant adds prompt-based multi-interest learning method into the Base variant. Specifically, we fed the prompt-augmented user interaction embeddings~$\bm{H}_{u,t}^\mathcal{F}$ and $\bm{H}_{u,t}^\mathcal{G}$ into the multi-interest extractor and aggregator, respectively. 
\item \textbf{+Disp.} Compared to the Base variant, this variant adopts the proposed centrality-dispersion based multi-interest extractor to derive the user multi-interest embeddings, i.e., $\bm{V}_{u, t} = \bm{M}_{u,t} +\lambda\bm{\Sigma}_{u,t}$.
\end{itemize}

The experimental results of the ablation study are displayed in Table~\ref{table_ablation_study}. Without losing generality, we reported the Recall@5, Recall@10, NDCG@5, and NDCG@10 of the methods. 
As can be seen, firstly, the variant +Prompt outperforms the variant Base, when incorporating the prompt embeddings. This demonstrates that it is helpful to add specific prompts for the user interaction embeddings to guide their learning objectives in the multi-interest extractor and aggregator.
Secondly, the variant \mbox{+Disp.} outperforms the variant Base, which proves that it is beneficial to learn the user multi-interest embedding from both the centrality and dispersion of the user interactions. 

\begin{table}[t]\small
\caption{Deployments of the PoMRec in existing multi-interest learning methods.}
\centering
\label{table_extension}
\renewcommand{\arraystretch}{1.3}
\setlength{\tabcolsep}{2mm}{
\begin{tabular}{crcccc}
\hline
\multirow{2}{*}{Dataset} & \multirow{2}{*}{Variant} & \multicolumn{2}{c}{Recall@N} & \multicolumn{2}{c}{NDCG@N} \\ \cmidrule(lr){3-4} \cmidrule(lr){5-6}
 & & N=5 & N=10 & N=5 & N=10 \\ \hline
 \multirow{6}{*}{ML-1M} 
& MINER & 0.2758 & 0.3901 & 0.1878 & 0.2246\\ 
& PoM-MINER &  \textbf{0.2975} &\textbf{0.4147}&\textbf{ 0.2066} & \textbf{0.2441}\\ \cline{3-6}
& Impro. & 7.87\%			
 & 6.31\% & 10.01\%& 8.68\%\\ \cline{2-6} 
& TiMiRec & 0.3091 & 0.4310 & 0.2136 & 0.2529\\  
& PoM-TiMiRec &  \textbf{0.3222} & \textbf{0.4344}& \textbf{0.2228} &\textbf{0.2580}\\ \cline{3-6} 		
& Impro. & 4.24\% & 0.79\% & 4.31\%& 2.02\%\\ \hline  
& MINER & 0.2467 &0.3398& 0.1702 & 0.2048 \\ 
& PoM-MINER &  \textbf{0.2652} &\textbf{0.3596}& \textbf{0.1870} &\textbf{0.2176}\\ \cline{3-6}
Movie& Impro. & 7.50\%		
 & 5.83\% & 9.87\%	& 6.25\%\\ \cline{2-6} 
\& TV& TiMiRec & 0.2506 &0.3412& 0.1755 &0.2048\\ 
& PoM-TiMiRec &  \textbf{0.2735} &\textbf{0.3674}& \textbf{0.1933} &\textbf{0.2236}\\ \cline{3-6} 
& Impro. & 9.14\%& 7.68\%& 10.14\%& 9.18\%\\ 
\hline 
\end{tabular}}
\end{table}

\subsection{Deployment on Other Methods (RQ3)}
The two proposals can also be deployed to other multi-interest learning methods that involve the multi-interest extractor and aggregator. In this part, we deployed our method to two state-of-the-art multi-interest learning method, i.e., MINER~\cite{LiZBCSDJL22} and TiMiRec~\cite{WangWLGM00FDM22}.
In particular, we directly adopted the model architectures of the multi-interest extractor and aggregator in MINER and TiMiRec. Differently, we added the learnable prompt embeddings into the inputs of the two modules and learned the user multi-interest embeddings with both the mean and variance embeddings of user interests, as same as our proposed PoMRec. The deployments of PoMRec on MINER and TiMiRec are termed as PoM-MINER and PoM-TiMiRec, respectively. Without losing generality, we reported the Recall@5 and NDCG@5 of the methods in ML-1M and Movie \& TV dataset in Table~\ref{table_extension}. 

As can be seen, the deployments PoM-MINER and PoM-TiMiRec outperform their corresponding methods MINER
and TiMiRec in both datasets, which demonstrates the effectiveness and applicability of the proposed method.
Notably, as aforementioned, the original TiMiRec assigns two embeddings for each item, i.e., one for the multi-interest extractor and the other one for the multi-interest aggregator. Nevertheless, our method aims to adapt the same inputs to different modules. Therefore, in PoM-TiMiRec, we only assigned a single embedding for each item. Intuitively, this adaptation could hurt the performance of PoM-TiMiRec. However, equipped the our proposed method, PoM-TiMiRec still works better than TiMiRec, which further proves our effectiveness.

\begin{figure*}[!t]
\centering
 \includegraphics[width=\linewidth]{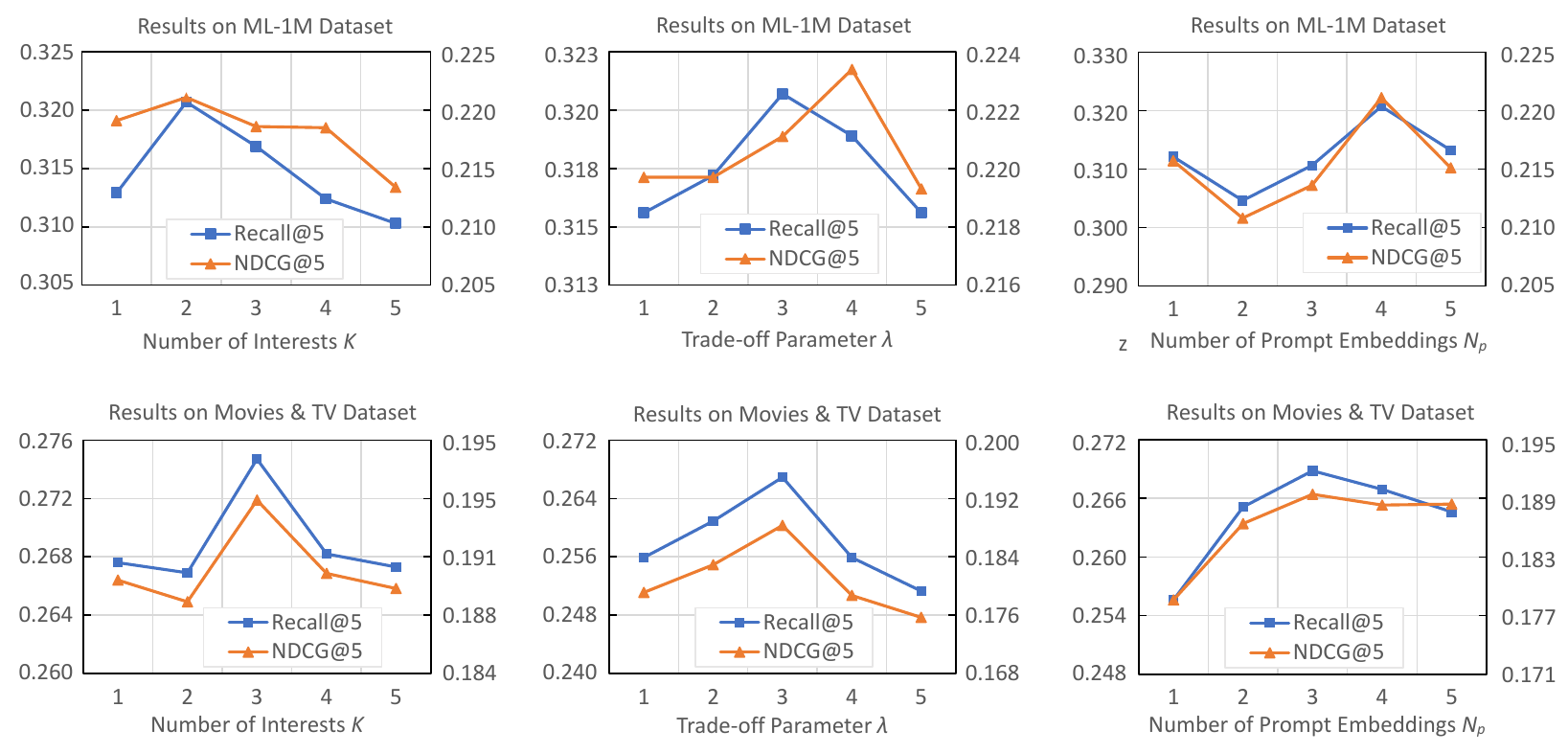}
\caption{Performance of the proposed PoMRec with respect to the different hyper-parameters in ML-1M and Movie \& TV datasets. The left vertical axis refers to the Recall@5, while the right vertical axis refers to the NDCG@5.}
\label{fig_parameters}
\end{figure*}

\subsection{Hyper-parameter Discussion (RQ4)}
\label{section_parameter}
In this subsection, we evaluated the following three key hyper-parameters: the number of user interests~$K$ defined in Eqn.~(\ref{eqn_interest_embeddings}), the trade-off parameter~$\lambda$ between the centrality and dispersion of user interactions during learning the user multi-interests defined in Eqn.~(\ref{eqn_new_interest_embedding}), and the number of the prompt embeddings~$N_p$ defined in Eqn.~(\ref{eqn_prompt_embeddings}).

\subsubsection{The Number of User Interests~$K$}
\label{section_K_discussion}
A larger number of user interests $K$ indicates that more interests the user has.
In this experiment, we fixed the other hyper-parameters to their default values and tested $K$ from 1 to 5 with a stride of $1$. Without losing generality, we reported the Recall@$5$ and NDCG@$5$ on both datasets in Figure~\ref{fig_parameters}. 
The results show that, for both datasets, as the number of user interests~$K$ increases from 1 to 5, the model performance first increases until it achieves the best performance with the most suitable $K$, and then decreases. This suggests that the more complex multi-interest embeddings may not lead to better performance. In fact, the overly complex multi-interest embeddings can complicate the training and may degrade the recommendation performance.
Furthermore, we observed that the most suitable $K$ for ML-1M dataset is $2$, while that for \mbox{Movies \& TV} dataset is $3$. This may be because that the interests of users in ML-1M are relatively more concentrated, while those in the other dataset are relatively more diverse.

\subsubsection{The Trade-off Parameter~$\lambda$}
\label{section_lambda_discussion}
The trade-off parameter~$\lambda$ balances the importance of the centrality and dispersion of user interactions during learning the learning of user multi-interests. In this experiment, we fixed the other hyper-parameters to their default values and tested $\lambda$ from $\{1,2,3,4,5\}$. The Recall@$5$ and NDCG@$5$ results are shown in Figure~\ref{fig_parameters}. As can be seen, the model performance first increases until achieving the best performance and then decreases. This demonstrates that involving the dispersion of the user interactions appropriately does benefit the user multi-interest learning. The most suitable $\lambda$ for ML-1M dataset is $4$, while that for \mbox{Movies \& TV} dataset is $3$. This may be because that the user interactions in the ML-1M dataset are more dispersed than those in \mbox{Movies \& TV} dataset, indicting that the dispersion of the user interactions plays a more important role in ML-1M.

\subsubsection{The Number of Prompt Embeddings~$N_p$}
The hyper-parameter~$N_p$ controls the number of the prompt embeddings added to the inputs of user interactions. A larger~$N_p$ increases the adaptive ability of the inputs to different learning objectives in the multi-interest extractor and aggregator. In this experiment, we fixed the other hyper-parameters to their default values and tested $N_p$ from $\{1,2,3,4,5\}$. The Recall@$5$ and NDCG@$5$ results are shown in Figure~\ref{fig_parameters}. From Figure~\ref{fig_parameters}, we can see that the performance first raises along with the number of the prompt embeddings~$N_p$ increasing in the two datasets. This demonstrates that it is helpful to insert appropriate prompts to make the inputs adaptive to the different learning objectives. However, the model performance decreases when $N_p$ becomes overly large. This may be because that the multi-interest extractor and aggregator have certain connections despite having different learning objectives for the inputs. For example, the learned user multi-interest embeddings of extractor could help learn their weights in the aggregator. 
Therefore, excessive prompt embeddings can significantly increase the difference between the extractor and aggregator, while lessening their useful connections.

\begin{figure*}[!t]
\centering
\includegraphics[width=\linewidth]{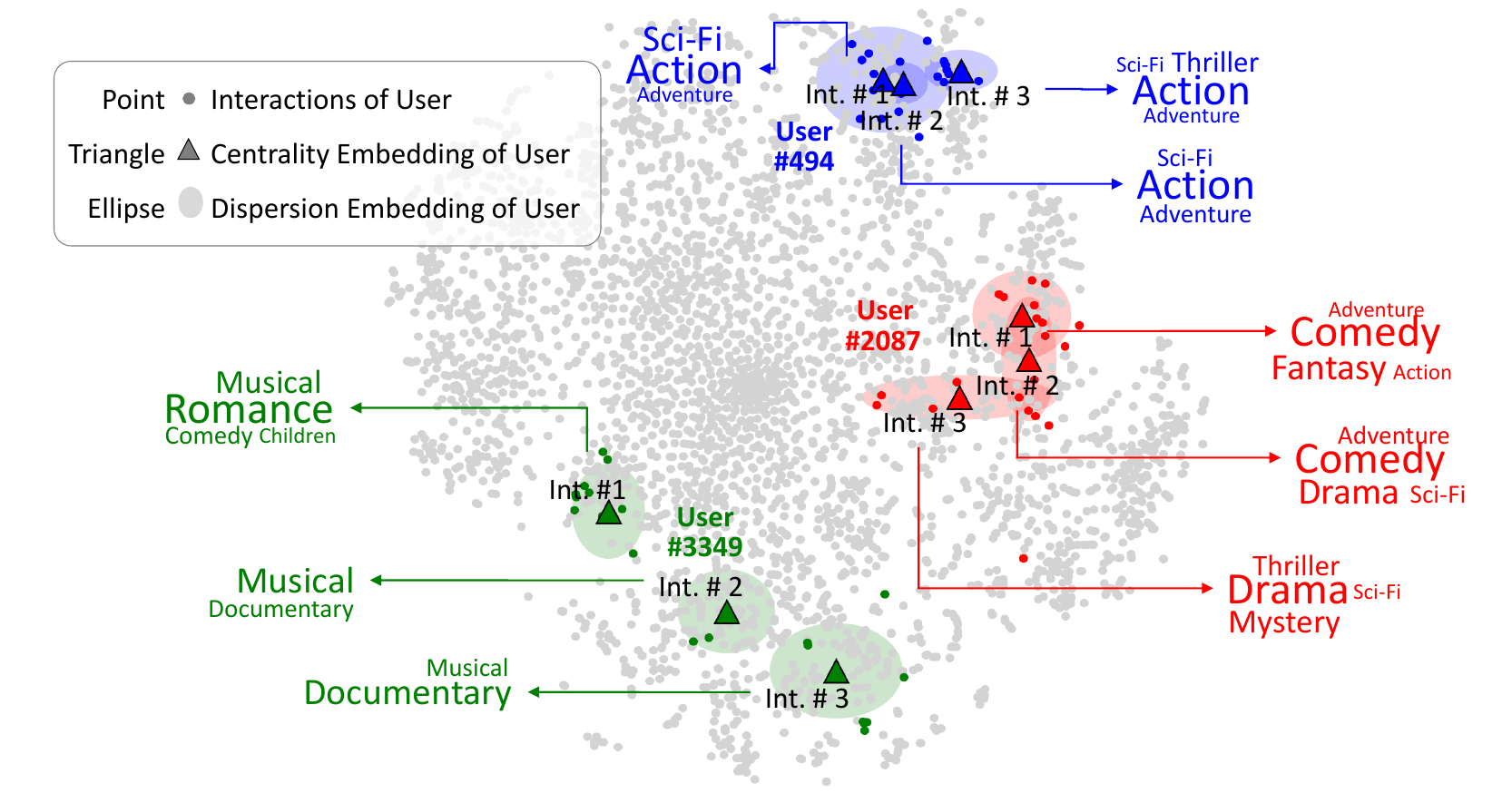}
\caption{Visualization of the learned user multi-interests in ML-1M dataset with the tool of t-SNE. The grey points represent the embeddings of all items in the dataset. The interactions of three users are highlighted with different colors. We visualize the learned user multi-interests with triangles (representing the centrality) and ellipses (representing the dispersion). To derive a deep understanding of the user multi-interests, we annotate the word cloud of user each interest generated with the item genres.}
\label{fig_visualization}
\end{figure*}

\subsection{Visualization of The User Multi-interests (RQ5)}
To intuitively show the necessity of the centrality and dispersion of the user interactions to capture the user multi-interests, in this subsection, we visualized three examples of the learned user multi-interest embeddings. In particular, we visualized the learned item embeddings of the ML-1M dataset with the tool of t-SNE~\cite{Maaten14} in Figure~\ref{fig_visualization}. Specifically, we highlighted the historical interactions of three random users, i.e., User~$\#494$, User~$\#2087$, and User~$\#3349$ in blue, red, and green, respectively. 
To depict the learned user multi-interest embeddings, for the $k$-th interest of the user~$u$, we utilized the t-SNE tool to map the learned mean vector~$\bm{\mu}_{u, t}^{(k)}  \in \mathbb{R}^{d}$ and variance vector~$\bm{\Sigma}_{u, t}^{(k)}  \in \mathbb{R}^{d}$ into a two-dimensional vector, respectively. Then the triangle is drawn with the mapped mean vector, and based on that the ellipse is drawn with the mapped variance vector as the radius. Additionally, to obtain a deeper insight of the learned user multi-interests, we annotated the genres of the items belonging to each user interest by the word clouds, which help us understand the semantics of the three users' multi-interests. The larger the font size, the more frequently the genre occurs in the interest. 
From Figure~\ref{fig_visualization}, we have the following observations:
\begin{itemize}[leftmargin=5mm]
\item The user interests are indeed diverse and multi-faceted, making it helpful to learn multiple interest embeddings for each user to capture their multi-interests. As can be seen, the interactions of green User~$\#3349$ can be clearly divided into three groups. Checking the word clouds of item genres in the groups, we found that each group of items has clearly common genres, while different groups have different genres. Specifically, the most frequent genres of the Interest~$\#1, \#2$, and $\#3$ of the green User~$\#3349$ are Romance, Musical, and Documentary, respectively.
\item Although we have learned a fixed number of interests for each user, the proposed PoMRec can deal with cases where a user has fewer interests than the fixed number. For example, as can be seen in Figure~\ref{fig_visualization}, the interactions of the blue User~$\#494$ are mostly gathered in one area. In such case, the learned mean embeddings, i.e., blue triangles, of the user multi-interests are close to each other, and the variance embeddings, i.e., blue ellipses, are almost overlapping. In addition, checking the word clouds of genres in interests of the blue User~$\#494$, we found that these three interests consist of mostly movies with genres of Action. Accordingly, we can infer that this user essentially has only one prominent interest in movies, i.e., the action movies. In this case, it is reasonable that the learned user multi-interests of PoMRec are close to each other.
\item The dispersion of the user interactions can complement the centrality, which helps better describe the user multi-interests. For example, as for the red User~$\#2087$ in Figure~\ref{fig_visualization}, the interactions are divided into three groups. Interestingly, the interactions belonging to Interest~$\#3$ are dispersed in the horizontal axis but concentrated along the vertical axis. In such case, the mean embedding of the user interactions, i.e., the third red triangle, cannot represent the two interactions in the left. The learned variance embedding, i.e., the third red ellipse, captures the dispersion of the user interactions in Interest~$\#3$ and covers the left interactions of the user. 
\end{itemize}

\section{Conclusion and Future Work}
\label{section_conclusion_and_future}
In this paper, we propose a prompt-based multi-interest learning method (PoMRec) for the sequential recommendation. In particular, we insert specific prompts into the user interactions to make them adaptive to the different learning objectives of the multi-interest extractor and aggregator. Moreover, we learn user multi-interest embeddings with not only the centrality of the user interactions but also their dispersion, which could comprehensively capture the user interests. Extensive experiments on three public datasets have demonstrated the effectiveness of the proposed PoMRec. 
In particular, we have found that although the multi-interest extractor and aggregator have their own learning objectives, they still share certain connections. Therefore, excessive specific prompts can introduce much differences between the two modules, while reducing their connections and hence negatively impacting the model performance.  
Nevertheless, the current PoMRec only utilizes the user interaction data, while overlooking the valuable information in item multimodal features. In the future, we plan to devise a multimodal multi-interest learning method to enhance the recommendation. 

\ifCLASSOPTIONcompsoc
\else
\fi


\bibliographystyle{IEEEtran}
\bibliography{ref.bib}

\begin{IEEEbiography}[{\includegraphics[width=1in,height=1.25in,clip,keepaspectratio]{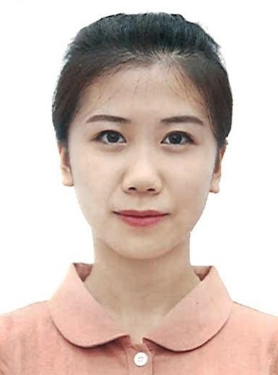}}]{Xue Dong} received the Ph.D. degree from the School of Software, Shandong University, China in 2023. She is currently a research fellow with the Tsinghua University, China. Her research interests contain multimedia computing, multimodal recommendation and retrieval. She has published several papers in the top venues, such as ACM SIGIR, MM, TOIS, IEEE TNNLS.
\end{IEEEbiography}

\begin{IEEEbiography}[{\includegraphics[width=1in,height=1.25in,clip,keepaspectratio]{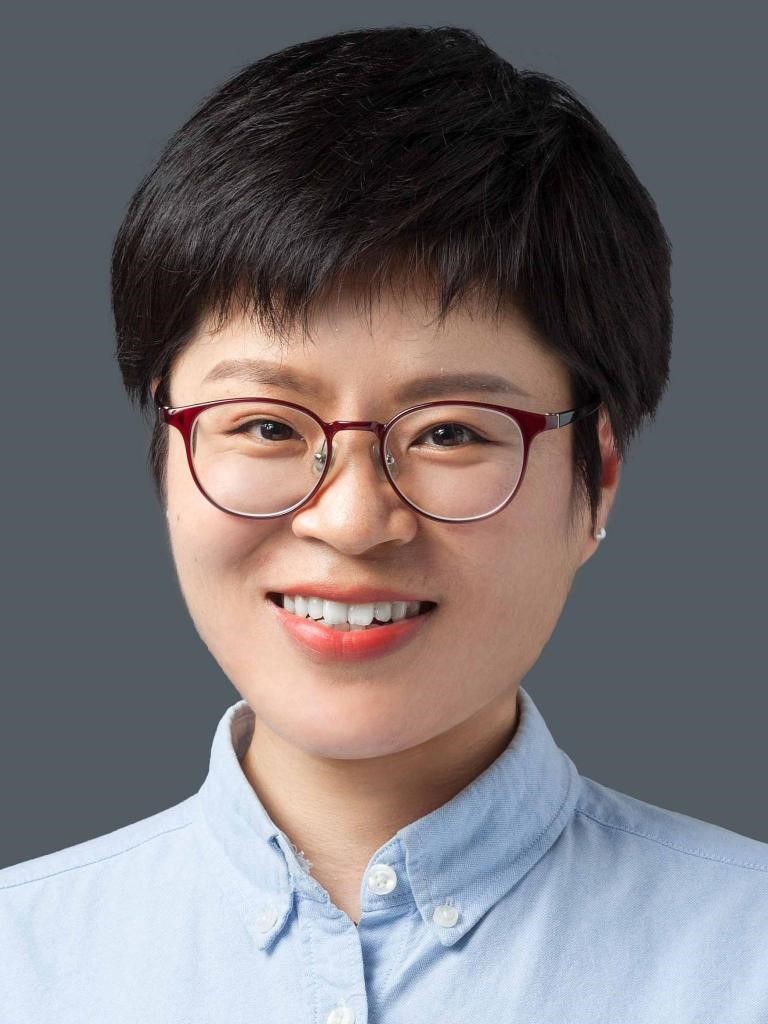}}]{Xuemeng Song} received the B.E. degree from the University of Science and Technology of China, in 2012, and the Ph.D. degree from the School of Computing, National University of Singapore, in 2016. She is currently an Associate Professor with Shandong University, China. She has published several papers in the top venues, such as ACM SIGIR, MM, and TOIS. Her research interests include information retrieval and social network analysis. She has served as a reviewer for many top conferences and journals.
\end{IEEEbiography}


\begin{IEEEbiography}[{\includegraphics[width=1in,height=1.25in,clip,keepaspectratio]{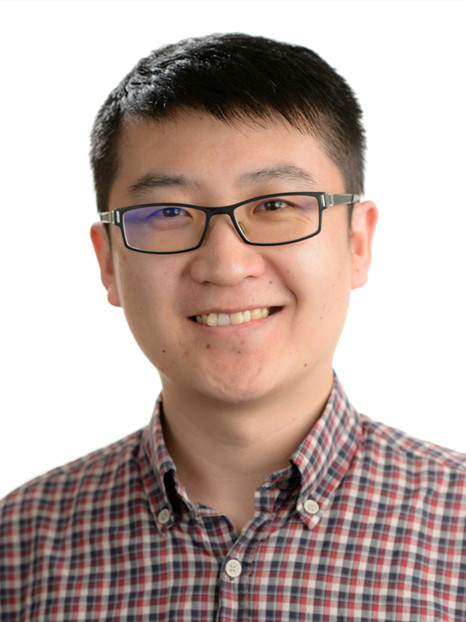}}]{Tongliang Liu} is the Director of Sydney AI Centre at the University of Sydney. He is also heading the Trustworthy Machine Learning Laboratory. He is broadly interested in the fields of trustworthy machine learning and its interdisciplinary applications, with a particular emphasis on learning with noisy labels, adversarial learning, transfer learning, unsupervised learning, and statistical deep learning theory. He has authored and co-authored more than 100 research articles including ICML, NeurIPS, ICLR, CVPR, ICCV, ECCV, AAAI, IJCAI, KDD, IEEE T-PAMI, T-NNLS, and T-IP. He is/was a (senior-) meta reviewer for many conferences, such as ICML, NeurIPS, ICLR, UAI, AAAI, IJCAI, and KDD. He is a recipient of Discovery Early Career Researcher Award (DECRA) from Australian Research Council (ARC) and was named in the Early Achievers Leaderboard of Engineering and Computer Science by The Australian in 2020.
\end{IEEEbiography}

\begin{IEEEbiography}[{\includegraphics[width=1in,height=1.25in,clip,keepaspectratio]{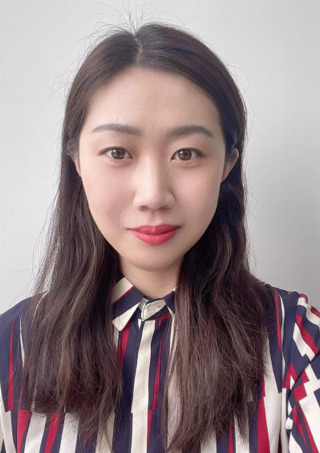}}]{Weili Guan} received the master degree from National University of Singapore. After that, she joined Hewlett Packard Enterprise in Singapore as a Software Engineer and worked there for around five years.  She is currently a PhD student with the Faculty of Information Technology, Monash University (Clayton Campus), Australia. Her research interests are multimedia computing and information retrieval. She has published many papers at the first-tier conferences and journals, like ACM MM, SIGIR, and IEEE TIP.
\end{IEEEbiography}


\end{document}